\documentclass[aps,superscriptaddress,floats,showpacs]{revtex4}

\usepackage{graphicx}
\usepackage{subfigure}
\begin{document}

\title{Dispersion of tracer particles in a compressible flow}
\author{John~R.~Cressman}
\affiliation{Department of Physics and Astronomy, University of
Pittsburgh, Pittsburgh, PA 15260, USA}
\author{Walter~I.~Goldburg}
\affiliation{Department of Physics and Astronomy, University of
Pittsburgh, Pittsburgh, PA 15260, USA}
\author{J\"org Schumacher}
\affiliation{Fachbereich Physik, Philipps-Universit\"at, D-35032
Marburg, Germany}
\date{\today}

\begin{abstract}
The turbulent diffusion of Lagrangian tracer particles has been studied
in a flow on the surface of a large tank of water and in computer simulations.
The effect of flow compressibility is captured in images of particle
fields.  The velocity field of floating particles has a divergence, whose
probability density function shows exponential tails.  Also studied is the motion
of pairs and triplets of particles.  The mean square separation $\langle
\Delta(t)^2 \rangle$ is fitted to the scaling form $\langle \Delta(t)^2 \rangle
\propto t^{\alpha}$, and in contrast with the Richardson-Kolmogorov prediction,
an extended range with a reduced scaling exponent of $\alpha=1.65\pm 0.1$ 
is found.
Clustering is also manifest in strongly deformed triangles spanned within
triplets of tracers.
\end{abstract}
\maketitle

{\em Introduction}.--- 
Many processes in nature depend on the efficient
diffusion in a turbulent medium such as spreading of drifters on the ocean
surface \cite{davis1991,lacasce2003} or of aerosol particles that can act as
nuclei for the formation of rain droplets in the atmosphere
\cite{facchini1999,shaw2003}.  If the flow is compressible (or the particles
have inertia), the following can occur:  stirring an initially uniform
distribution of particles can induce them to coagulate, as can be seen in Fig.
1 (left).  This photo was taken a fraction of a second after floating particles were
uniformly dispersed on the surface of a large tank of water (lateral dimensions
1 m $\times$ 1 m) that was continuously sustained in a turbulent state by bulk
driving.  The particles appear in white in this figure.  After a few seconds
they accumulate in narrow ridges with voids in between, forming a lace-like
network.

The origin of the coagulation is apparent in the right panel of Fig.~1. The
incompressible fluid below the surface S is shown to be moving upward near 
point U
and downward near point D.  The floaters cannot follow the bulk motion; rather
they accumulate at points like D and flee points like U.  The floating particles
move in the $x-y$ plane, which will be taken at $z=0$.  They simply sample the
horizontal components of the flow velocity, $v_{x}(x,y,0,t)$ and
$v_{y}(x,y,0,t)$, of the incompressible fluid below them.  Since water is
incompressible, the divergence of the velocity field is zero at all points on
the surface and below it, so that $\partial v_{x}(x,y,z, t)/\partial x +\partial
v_{y}(x,y,z,t)/\partial y=-\partial v_{z}(x ,y,z,t)/\partial z$, which is
non-zero and, in fact, large.  Thus the floaters form a two-dimensional
compressible subsystem, even though the flow velocity is negligible compared to
the speed of sound and their behaviour is unrelated to
two-dimensional turbulence \cite{goldburg}.

\begin{figure}
\begin{center}
\includegraphics[width=6cm,height=7cm]{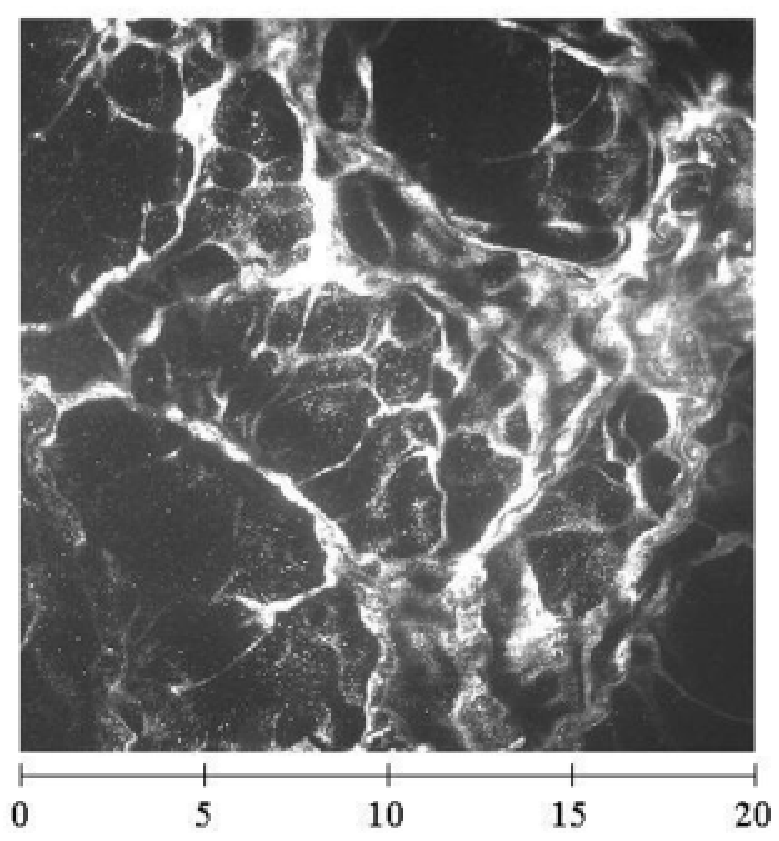}
\includegraphics[width=5cm,height=5cm]{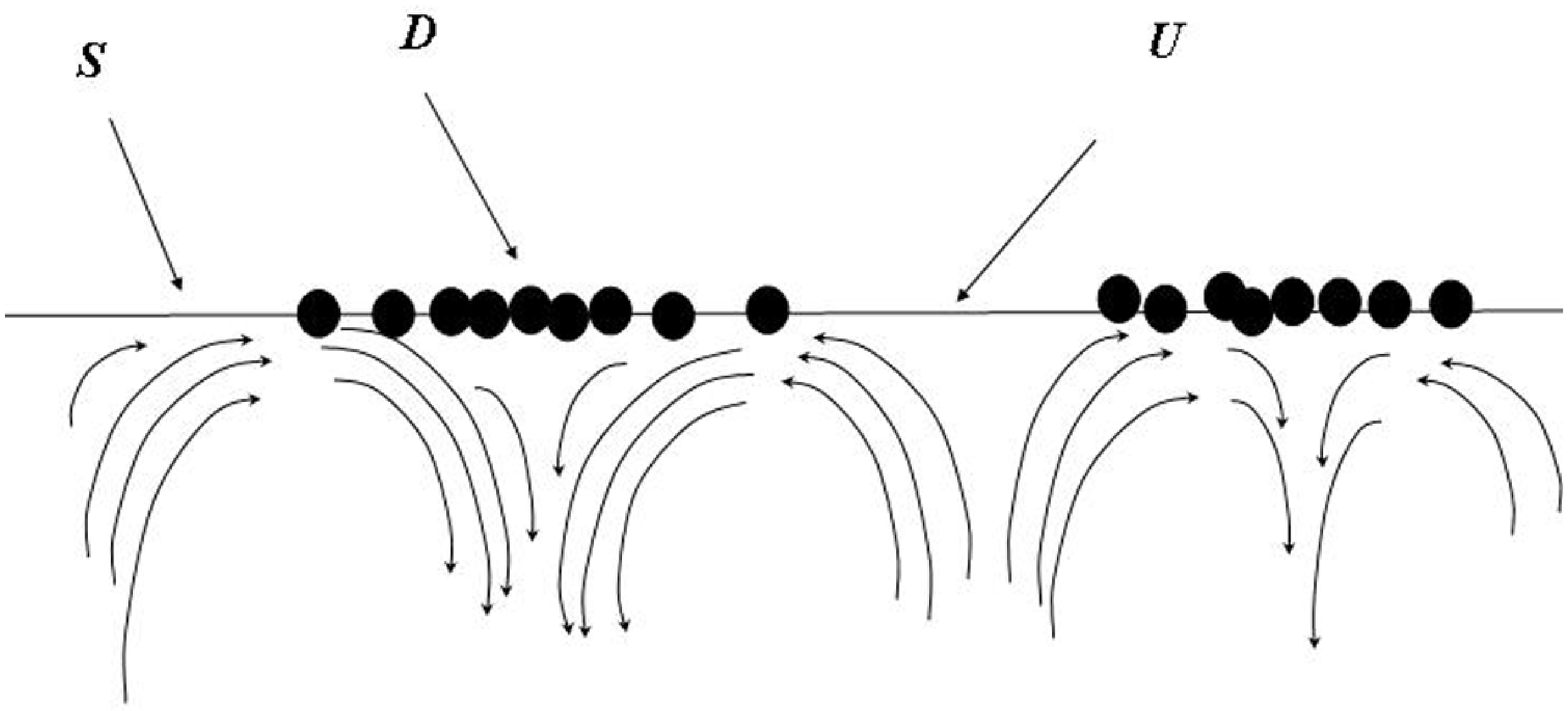}
\end{center}
\caption{Left: Image of a cloud of particles on the surface of a turbulent body of
water.  The particles which are 10 $\mu$m in diameter appear as white and were
initially dispersed uniformly over the surface.  This image was captured 100 ms
later.  The scale below the figure is in cm. Right:  Sketch
illustrating the origin of the clustering effect at the surface S.}
\end{figure}

Recently, compressible turbulence has come under theoretical scrutiny
(see \cite{falkovich} for a review).
Modelling the flow as Gaussian and delta-correlated in time, one can show for
both spatially smooth \cite{chertkov1998} and spatially rough flows
\cite{gawedzki2000}, that a transition from a regime of weak to strong
compressibility occurs at some threshold value of the dimensionless ratio,
defined as
\begin{equation}
{\cal C} \equiv  \langle(\partial v_i/\partial x_i)^{2}\rangle/\langle(\partial
v_i/\partial x_j)^2\rangle\,,
\end{equation}
with $i,j=1,2$. Clearly, for any ${\cal C}\ne 0$ tracers will cluster, 
but the eventual
particle distribution will depend on whether ${\cal C}$ is 
above or below the threshold.
Assumptions contained in those models make it inapplicable to the present
experimental situation, where the flow is non-Gaussian
and has a finite time correlation.  Because the idealizations in those
models are not realized in the experiment, we do not know if such a transition
from weak 
to strong compressibility exists in our experiment.  
On the other hand, a computer simulation (DNS) that roughly matches the flow
conditions of this experiment (incompressibility of the three-dimensional
velocity field at all $z \leq 0$, finite correlation time, and a free-slip
boundary condition at the flat surface) are in remarkably good agreement with
the first experiments \cite{goldburg2001,eckhardt2001}. Herein, we summarize 
and extend recent experiments using freshly cleaned water surfaces 
\cite{goldburg2003}. 

Our study focuses on two points.  The first one is the relative motion
of the floaters, with special emphasis on so-called Richardson diffusion
\cite{richardson1926}, which refers to the time dependence of the mean square
particle separation $\langle \Delta (t)^2 \rangle\sim t^{\alpha}$ with 
$\alpha=3$.  Scaling close to the
Richardson prediction was found in experiments \cite{tabeling99} as well as in
high-resolution numerical simulations \cite{celani00} in the inverse cascade of
two-dimensional incompressible turbulence.  Intermittency effects if present
seem to modify this exponent only slightly \cite{boffetta2002}.  Dimensional
arguments would seem to be inapplicable for the floaters, since they can
exchange kinetic energy and mean square vorticity with the fluid below them.
Since a theory for their motion has yet to be developed, one must again turn to
computer simulations, which have been recently carried out
\cite{schumacher2002}.

Secondly, first studies the competing effects of coherent shear and 
random motion in the surface flow are presented by analyzing 
the evolution of shapes formed by three particles tracked
simultaneously \cite{pumir2000,castiglione2001}.  By neglecting the center of
mass, the relative position of three points can be conveniently expressed by the
following two vectors, ${\bf a}_1=({\bf x}_2-{\bf x}_1)/\sqrt{2}$ and ${\bf
a}_2=(2{\bf x}_3-{\bf x}_1-{\bf x}_2)/\sqrt{6}$ \cite{castiglione2001}.  The
distortion of a triangle is measured by the ratio $I_2=g_2^2/R^2$ which relates
its radius of gyration, $R=\sqrt{a_1^2+a_2^2}$, to the smaller half-axis, $g_2$,
of the smallest ellipse that covers the tracer triangle.  From
\cite{castiglione2001} we get for two dimensions
\begin{equation}
I_2=\frac{1}{2}\left[ 1-\sqrt{1-\frac{4}{R^4}({\bf a}_1\times {\bf
a}_2)^2}\,\right]\,.
\label{I2}
\end{equation}
We will present measurements
of the evolution of $\langle I_2 \rangle$ as a function of time and propose a
qualitative explanation based on the compressibility of the flow.

{\em Experimental setup}.--- 
The square tank in which the measurements were
made was filled to a depth $Z$, which was typically 30 cm, and the turbulence
was generated by a variable speed 8 hp pump that circulated the water through a
series of pipes capped with rotating jets located well below the surface.  The
turbulent intensity was limited to keep the amplitude of the surface waves below
1 mm.  Images from a fast overhead camera, which typically operated at 300
frames/s, were captured in a computer and analyzed with in-house software.  A 5
W diode pumped solid state laser was used to produce a sheet of light
illuminating only the surface.

Measurements of the velocity well below
the water surface yielded a second order longitudinal structure function 
$D_2(r) \propto r^{0.65 \pm 0.05}$ in the interval 1.2 mm $< r < 3.5$cm.
This is close to the Kolmogorov value of 2/3.
On the surface, the outer scale of the turbulence $r_0$ was 3.2 cm.  Essential
to the success of the experiment was the continuous vacuuming of the surface by
a small auxiliary pump connected to a vertical pipe that comes up to the
surface.  Without this cleaning, which was already reported in 
\cite{goldburg2003}, the particles will begin to interact as their
density becomes large, destroying the clustering seen in Fig.~1. 
The floating
particles used in velocity measurements on the surface were hollow glass spheres
with radius $a$ = 200 $\mu$m and highly buoyant particles of specific gravity
0.25 with $a$ =50 $\mu$m.  The Stokes time, $\tau_S=2 a^2/(9\nu)$, was roughly 1
$\mu$s, so that tracers easily follow the flow beneath them ($\nu$ is the
kinematic viscosity).  Other particles, such as mushrooms spores and talc were
also used successfully.

\begin{figure}
\begin{center}
\includegraphics[width=11cm]{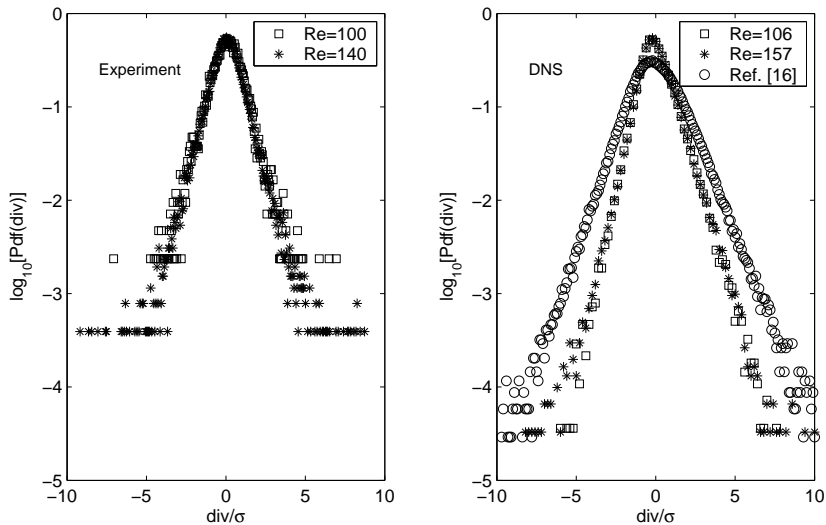}
\end{center}
\caption{Probability density function (PDF) of the surface flow divergence field
with $\sigma=\langle({\bf\nabla\cdot v})^2\rangle^{1/2}$.  Left panel shows the
experimental findings for two Taylor microscale Reynolds numbers of $Re=100$ and
140, respectively.  Right panel shows the simulation results for three different
values.  Runs at $Re=106$ and $157$ are for driving by a large-scale shear.
Data from \cite{schumacher2002} were obtained with isotropic 
volume forcing at a Taylor Reynolds number of $Re=145$.}
\end{figure}

{\em Surface flow divergence}.--- 
By tracking the motion of the small surface
particles over a short interval of time, one can map out the velocity
$v_{x}(x,y,0,t), v_{y}(x,y,0,t)$ of the floaters and extract the local surface
divergence, $\nabla_2 \cdot {\bf v} = \partial v_{x}(x,y,t)/\partial x +\partial
v_{y}(x,y,t)/\partial y$, a quantity that will fluctuate in time and space.  The
two-dimensional divergence $\nabla_2 \cdot {\bf v}$ is therefore a random
variable, whose probability density function (PDF) $P_{\nabla_2{\bf\cdot v}}$ is
plotted in Fig.~2.  Those measurements were made at Taylor microscale Reynolds
numbers $Re$ of 100 and 140 (left panel).  
The relatively large size of the seed particles
made it impossible to directly measure velocity differences on a scale $r$ less
then the Kolmogorov dissipation scale, $\eta$, which is $\approx$ 0.1 mm.  Therefore, the
quoted Taylor microscale was determined from gradients measured over a few
millimeters.

New experimental data were extracted from an ensemble of measurements made at
many instants of time and many points $(x, y)$ in the surface.  Note that the
mean value of $\nabla_2{\bf\cdot v}$ is close to zero and that this function has
exponential tails out to almost four decades in its standard deviation,
$\sigma_{{\nabla_2}\cdot{\bf v}}$.  Computer simulations of
$P_{\nabla_2{\bf\cdot v}}$ for isotropic forcing \cite{schumacher2002} 
as well as large-scale
shear likewise show this exponential behaviour (see the right panel of Fig.~2).
Exponential behaviour for the PDF can be expected, because the divergence is
composed of velocity derivatives which are known to yield exponential tails in
two- and three-dimensions.  Note that the slope of the decay of the tails is
effected by the details of volume forcing and seems thus to be non-universal.
In both the computer simulations and the laboratory experiments, it was found
that ${\cal C} \approx 0.5$ regardless of the intensity or geometry of the
forcing. 

{\em Pair dispersion}.---
An important characteristic of turbulence is the
ability of velocity fluctuations to disperse particles.  Figure~3 displays
relative diffusion $\langle \Delta^2(t) \rangle$ for both laboratory experiments
(circles) and computer simulations (triangles).  One
expects that $\alpha$ =3 for an incompressible turbulent flow and times
larger than the Lagrangian decorrelation time which is $\sim 10$
Kolmogorov times, $\tau_{\eta}$.
Instead, both the simulations and the experiments on the present flow show a
crossover from $\alpha \simeq$ 2 to an exponent in the range
1.65 to 1.8.  Although the simulations display a small range where the slope is
approximately 3.2, the most striking feature of the pair dispersion measurements
is a reduced value of the scaling exponent $\alpha$ in the inertial-range for both
measurements and the simulations.

The reduced $\alpha$ reflects a correlated motion of particles within a pair:
(i) pair particles remain close together while moving in a big cluster or pair
trajectories merge again after longer transient phases of some separation, i.e.
their contribution to the growth of $\langle \Delta^2(t) \rangle$ with time is
small.  (ii) on the other hand, decorrelation due to breaking of clusters by
newly upwelling fluid or simply large scale separation does appear. 
This break-up
introduces a randomizing element and causes a diffusive Brownian scaling.  While
the latter mechanism can be found in an incompressible system as well, the first
scenario, (i), might be related to the compressibility of the advecting flow. 
Note that one has to be 
careful with the interpretation of our results. 
For large compressibilities, trapping of trajectories can be predicted for
the Kraichnan flow analytically \cite{gawedzki2000}.
Recently, Sokolov performed numerical
particle dispersion experiments in a synthetic incompressible flow where the
exponent $\beta$ of the spatial scaling of the distance dependent correlation 
time of the quasi-Lagrangian pair velocity difference could be varied, 
$\tau(\Delta)\sim \Delta^{\beta}$ \cite{sokolov01}. For
$\beta$ increasing from 0 to 2/3 the exponent $\alpha$ of the pair dispersion
increased from 3/2 to 3. It is still unclear how the compressibility affects
this correlation time scaling although first attempts have been made
\cite{gawedzki03}. 
The present resolution of the data prohibits a detailed analysis 
of this issue for a Navier-Stokes flow.

\begin{figure}
\begin{center}
\includegraphics[width=7cm]{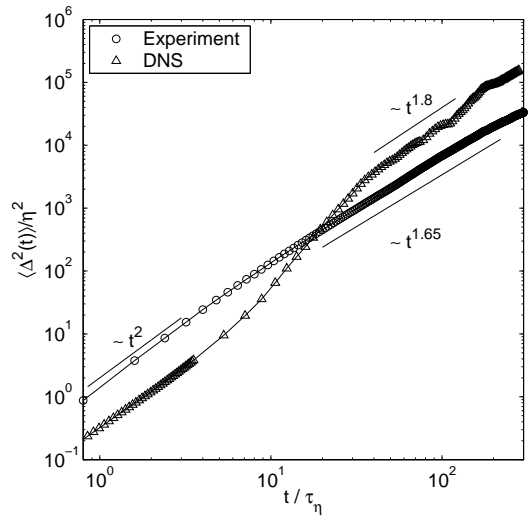}
\end{center}
\caption{Pair dispersion $\langle \Delta^2(t)\rangle/\eta^2$ vs.
$t/\tau_{\eta}$ for the surface flow experiment (open circles).  For a better
comparison we added the result of the numerical experiment (open triangles)
\cite{schumacher2002}.}
\end{figure}

Experimentally, $\tau_\eta$ is determined from ${\tau_\eta}^{-1} = \langle
\partial v_x/\partial x \rangle$ measured at $\eta$.  The value of $\tau_\eta$
was extrapolated from measurements at larger scales and was found to have a
maximum value of 1/80 sec.  The measurements of the surface velocity field,
$v_x(x,y,0,t), v_y(x,y,0,t)$, enable determination of the mean square pair
separation $\Delta^{2}(\Delta_0,t)$, where $\Delta_0$ is the pair separation at
$t= 0$.  It was not possible to accurately track the motion of true floaters
over large distances because of the slight vertical motion of the surface and
the time-varying illumination it produces.  Instead, the measured velocity
fields were used to propagate the motion of simulated particles where we checked
the robustness of the pair diffusion measurements with respect to variations of
the interpolation parameters.  Snapshots of the simulated particle
fields verify that the structures produced are qualitatively the same as for the
real particles seen in Fig.~1. The limited resolution of the data 
prevented an analysis of higher order moments of the particle density which
is another appropriate measure to characterize the clustering (for more details
see \cite{schumacher2002}).

{\em Shape distortion of tracer triplets}.---
In order to further probe the
effects of compressibility on flow structure, we investigated the distortion of
triangular configurations of floaters .  The experimental data for $I_2$ are in
the left upper box of Fig.~4 and the simulations are shown in the box to the
right.  In both cases the particles were initially placed to form equilateral
triangles, for which $I_2=1/2$ (i.e.  ${\bf a}_1\perp {\bf a}_2$ and $|{\bf
a}_1|=|{\bf a}_2|$).  The three curves are for triangles of different initial
side length.

Pumir et {\it al.}  \cite{pumir2000} argue that coherent shear is responsible
for flattening initially equilateral triangles, causing a decrease in $\langle
I_2 \rangle$.  After a sufficiently long time, random small-scale fluctuations
tend to drive $\langle I_2\rangle$ towards the Gaussian equilibrium value of
$I_{2 G}=(1-\pi/4)/2$.  In the two-dimensional incompressible case, the
experiments show that the Gaussian limit is achieved in roughly 20
$t/\tau_{\eta}$ \cite{castiglione2001}.  It was asserted in Ref.~\cite{pumir2000} that
coherent shear should dominate in the viscous subrange and give way to the
randomizing effects of turbulent mixing at larger scales.

The measurements and the simulations in Fig.~4 show a strong dip below the
asymptotic value of $I_{2 G}$ and an extended minimum that exists for triangles
initially having a radius of gyration significantly larger than the dissipative
scale.  This behaviour can be accounted for by strong clustering that is apparent
in Fig.~1.  The triplets are quickly compressed into quasi-colinear
configurations, making ${\bf a}_1$ almost parallel to ${\bf a}_2$ which forces
$I_2$, shown in Fig.~4, towards zero (see Eq.~(\ref{I2})).  In incompressible
turbulence, this effect is absent, and $I_2$ very soon attains its random
Gaussian value $I_{2G}$.  The laboratory data attain a value of
$I_2$ that is slightly larger than the Gaussian value.  The triangles in the
computer simulations show an even stronger tendency to remain flat and appear to
approach an equilibrium value smaller than $I_{2 G}$.  Self avoidance of the
finite-size particles and possibly the particle tracking resolution used in the
laboratory experiments force a limit on the minimum attainable value of $\langle
I_2 \rangle$ and presumably cause its later growth.  In the numerical study
(point) particles can become arbitrarily close and thus $\langle I_2 \rangle$
can remain at a small value well below Gaussian.

\begin{figure}
\begin{center}
\includegraphics[width=8cm,height=8cm]{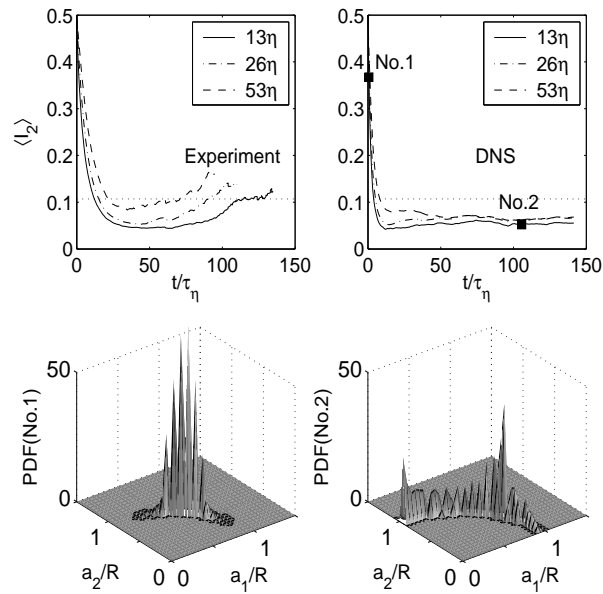}
\end{center}
\caption{Upper row:  Temporal evolution of $\langle I_2\rangle$.  Left panel is
for experiment at a Taylor Reynolds number of $Re=100$.  
Right panel is for numerical simulation at
$Re=145$.  The initial side lengths of triangles are given in the
legend.  The Gaussian asymptotic value of the two-dimensional incompressible
case of $I_{2 G}=(1-\pi/4)/2$ is indicated as a dotted line.  Lower row:
Probability density function $p(a_1/R,a_2/R)$ from the simulations for two
points in time and the initial triangle side length of $13 \eta$.  Both times
are indicated by filled squares in the upper right panel.  The support of the
PDF is the unit circle.  The initial equilateral case would correspond to a
delta-peak at 45 degrees.  This distribution spreads toward the cases where
$a_1\gg a_2$ and vice versa.}
\end{figure} 

The above results from the computer simulations suggest that the compressibility is acting to flatten
triangles, even when $R$ is much greater than $\eta$, contrary to incompressible
flows \cite{pumir2000}.  A dominant number of triangles remain distorted for
very long times.  This is illustrated in the two lower panels of Fig.~4, which
are plots of the joint PDF of the normalized vector magnitudes $|{\bf a}_1|/R$
and $|{\bf a}_2|/R$.  After a short time (lower left panel), the triangles are
only slightly distorted, which is reflected by weak scattering about the
equilateral value, $|{\bf a}_1|/R$ = $|{\bf a}_2|/R$.  The right panel shows that the
triangles have become strongly distorted at a sufficiently later time.
Here we observe two peaks ($a_1\gg a_2$ and vice versa).  It is notable that the
distribution function is confined within two sharply defined peaks rather than
spread widely through the $|{\bf a}_1|/R$- $|{\bf a}_2|/R$ plane.
Laboratory data pertaining to these distributions is not reported here.  However
the analogous PDF's have also been measured.  The finite size of the particles
blocks observation of the peak at $|{\bf a}_1|/R$ = 0 and $|{\bf a}_2|/R$=1,
apparent in the simulations, in the lower right panel of Fig.~4.

In summary, the relative motion of passive tracer particles in our {\em
compressible} flow is very different from tracer motion in incompressible fluid
turbulence.  The turbulence creates locally convergent and divergent regions of
particle density and inhibits Kolmogorov-Richardson pair dispersion.  
Likewise, the evolution of geometrical structures, revealed in the
motion of particle triplets, strongly differs from incompressible fluids.

{\em Acknowledgments}.--- Comments by B.~Eckhardt and J.~Davoudi are
acknowledged.  This work is supported by the National Science Foundation under
Grant No.  0201805 and the Deutsche Forschungsgemeinschaft.  Computations were
carried out on a Cray SV1ex at the John von Neumann-Institute for Computing at
the Forschungszentrum J\"ulich.


\end{document}